\documentclass[runningheads]{llncs}
\usepackage{amsmath}
\usepackage{amssymb}
\usepackage[english]{babel}
\usepackage{stackengine}
\usepackage{scalerel}
\usepackage[ruled]{algorithm2e}
\usepackage{graphicx}
\usepackage{svg}
\usepackage{subcaption}
\usepackage{amsmath}
\usepackage{url}
\usepackage[outercaption]{sidecap}    
\begin{document}
\title{Multiple-shooting adjoint method for whole-brain dynamic causal modeling 
}
%
%
\author{Juntang Zhuang, Nicha Dvornek, Sekhar Tatikonda, Xenophon Papademetris, Pamela Ventola, James Duncan}
\authorrunning{Juntang Zhuang, Nicha Dvornek, Sekhar Tatikonda, Xenophon Papademetris, Pamela Ventola, James Duncan}
%
\institute{Yale University}
\maketitle              
\begin{abstract}
Dynamic causal modeling (DCM) is a Bayesian framework to infer directed connections between compartments, and has been used to describe the interactions between underlying neural populations based on functional neuroimaging data. DCM is typically analyzed with the expectation-maximization (EM) algorithm.
However, because the inversion of a large-scale continuous system is difficult when noisy observations are present, DCM by EM is typically limited to a small number of compartments ($<10$). Another drawback with the current method is its complexity; when the forward model changes, the posterior mean changes, and we need to re-derive the algorithm for optimization.
In this project, we propose the Multiple-Shooting Adjoint (MSA) method to address these limitations. MSA uses the multiple-shooting method for parameter estimation in ordinary differential equations (ODEs) under noisy observations, and is suitable for large-scale systems such as whole-brain analysis in functional MRI (fMRI). Furthermore, MSA uses the adjoint method for accurate gradient estimation in the ODE; since the adjoint method is generic, MSA is a generic method for both linear and non-linear systems, and does not require re-derivation of the algorithm as in EM. We validate MSA in extensive experiments: 1) in toy examples with both linear and non-linear models, we show that MSA achieves better accuracy in parameter value estimation than EM; furthermore, MSA can be successfully applied to large systems with up to 100 compartments; and 2) using real fMRI data, we apply MSA to the estimation of the whole-brain effective connectome and show improved classification of autism spectrum disorder (ASD) vs. control compared to using the functional connectome. The package is provided \url{https://jzkay12.github.io/TorchDiffEqPack}
\end{abstract}

\section{Introduction}
Autism spectrum disorder (ASD) is a neurodevelopmental disorder that affects both social behavior and mental health \cite{nation2006patterns}. ASD is typically diagnosed with behavioral tests, and recently functional MRI (fMRI) has been applied to analyze the cause of ASD \cite{di2017enhancing}.
Connectome analysis in fMRI aims to elucidate neural connections in the brain and can be generally categorized into two types: the functional connectome (FC) \cite{van2010exploring} and the effective connectome (EC) \cite{friston2003dynamic}. The FC typically calculates the correlation between time-series of different regions-of-interest (ROIs) in the brain, which is typically robust and easy to compute; however, FC does not reveal the underlying dynamics.
EC models the directed influence between ROIs, 
and is widely used in analysis of EEG \cite{kiebel2008dynamic} and fMRI \cite{seghier2010identifying}. 

EC is typically estimated using dynamic causal modeling (DCM) \cite{friston2003dynamic}.
DCM can be viewed as a Bayesian framework for parameter estimation in a dynamical system represented by an ordinary different equation (ODE). A DCM model is typically optimized using the expectation-maximization (EM) algorithm \cite{moon1996expectation}. Despite its wide application and good theoretical properties, a drawback is we need to re-derive the algorithm when the forward model changes, which limits its application. Furthermore, current DCM can not handle large-scale systems, hence is unsuitable for whole-brain analysis. Recent works such as rDCM \cite{frassle2020regression}, spectral-DCM \cite{razi2017large} and sparse-DCM \cite{prando2020sparse} modify DCM for whole-brain analysis of resting-state fMRI, yet they are limited to a linear dynamical system and use the EM algorithm for optimization, hence cannot be used as off-the-shelf methods for different forward models.

In this project, we propose the Multiple-Shooting Adjoint (MSA) method for parameter estimation in DCM. Specifically, MSA uses the multiple-shooting method \cite{bock1984multiple} for robust fitting of an ODE, and uses the adjoint method \cite{pontryagin2018mathematical} for gradient estimation in the continuous case; after deriving the gradient, generic optimizers such as stochastic gradient descent (SGD) can be applied. 
Our contributions are: (1) MSA is implemented as an off-the-shelf method, and can be easily applied to generic non-linear cases by specifying the forward model without re-deriving the optimization algorithm. (2) In toy examples, we validate the accuracy of MSA in parameter estimation; we also validated its ability to handle large-scale systems. (3) We apply MSA in the whole-brain dynamic causal modeling for fMRI; in a classification task of ASD vs. control, EC estimated by MSA achieves better performance than FC.
\section{Methods}
We first introduce the notations and problem in Sec 2.1, then introduce mathematical methods in Sec2.3-2.4, and finally introduce DCM for fMRI in Sec2.5.
\subsection{Notations and formulation of problem}
\begin{figure}[t]
\begin{subfigure}{0.40\textwidth}
\includegraphics[width=\linewidth]{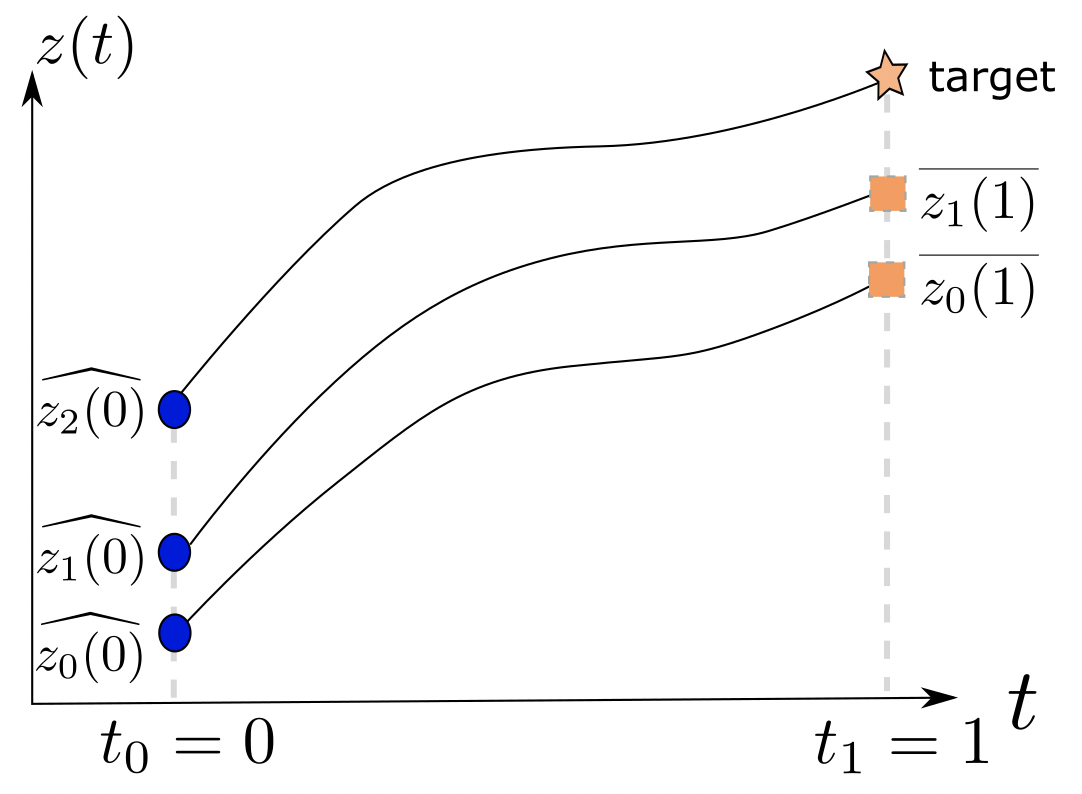}
\end{subfigure}
\hfill
\begin{subfigure}{0.54\textwidth}
\includegraphics[width=\linewidth]{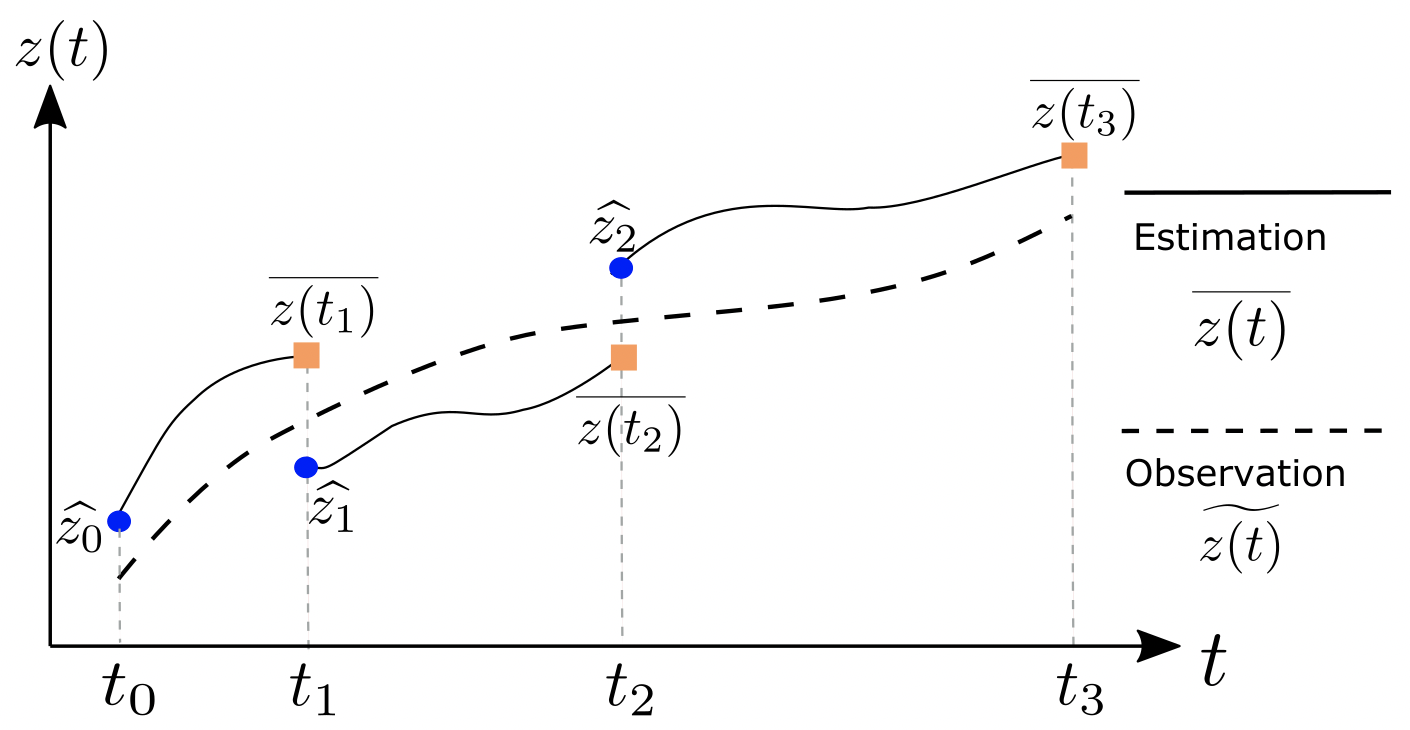}
\end{subfigure}
\caption{ \small{ Left: illustration of the shooting method. Right: illustration of the multiple-shooting method. Blue dots represent the guess of state at split time $t_i$.}}
\label{fig:shoot}
\end{figure}

We summarize notations here for the ease of reading, which correspond to Fig.~\ref{fig:shoot}.
\renewcommand\labelitemi{$\bullet$}
\begin{itemize}  
\item $z(t), \widetilde{z(t)}, \overline{z(t)}$: $z(t)$ is the true time-series, $\widetilde{z(t)}$ is the noisy observation, and $\overline{z(t)}$ is the estimation. If $p$ time-series are observed, then they are $p$-dimensional vectors for each time $t$.\vspace{1mm}
\item $(t_i,\widehat{z_i})_{i=0}^N$: $\{\widehat{z_i}\}_{i=0}^N$ are corresponding guesses of states at split time points $\{t_i\}_{i=0}^N$. See Fig.~\ref{fig:shoot}. $\widehat{z_i}$ are discrete points, while $\widetilde{z(t)}, z(t), \overline{z(t)}$ are trajectories. \vspace{1mm}
\item $f_\eta$: Hidden state $z(t)$ follows the ODE $\frac{d z}{dt}=f(z,t)$, $f$ is parameterized by $\eta$.\vspace{1mm}
\item $\theta$: $\theta=[\eta, z_0, ... z_N]$. We concatenate all optimizable parameters into one vector for the ease of notation, denoted as $\theta$. \vspace{1mm}
\item $\lambda(t)$: Lagrangian multiplier in the continuous case, used to derive the adjoint state equation.
\end{itemize}
The task of DCM can be viewed as a parameter estimation problem for a continuous dynamical system, and can be formulated as:
\begin{equation}
\label{eq:formulation}
    \operatorname*{argmin}_{\eta} \int \Big( \overline{z(\tau)} - \widetilde{z(\tau)} \Big)^2 d\tau \ \ \ s.t.\ \ \frac{d \overline{z(\tau)}}{d \tau} = f_\eta (\overline{z(\tau)},\tau)
\end{equation}
The goal is to estimate $\eta$ from observations $\widetilde{z}$.
In the following sections, we first briefly introduce the multiple-shooting method, which is related to the numerical solution of a continuous dynamical system; next, we introduce the adjoint state method, which efficiently determines the gradient for parameters in continuous dynamical systems; next, we introduce the proposed MSA method, which combines multiple-shooting and the adjoint state method, and can be applied with general forward models and gradient-based optimizers; finally, we introduce the DCM model, and demonstrate the application of MSA. 
\subsection{Multiple-shooting method}
The shooting method is commonly used to fit an ODE under noisy observations, which is crucial for parameter estimation in ODE. In this section, we first introduce the shooting method, then explain its variant, the multiple-shooting method, for long time-series.
\subsubsection{Shooting method} The shooting method typically reduces a boundary-value problem to an initial value problem \cite{hildebrand1987introduction}. An example is shown in Fig.~\ref{fig:shoot}: to find a correct initial condition (at $t_0=0$) that reaches the target (at $t_1=1$), the shooting algorithm first takes an initial guess (e.g. $\widehat{z_0(0)}$), then integrate the curve to reach point $(t_1, \overline{z_0(1))}$; the error term $target-z_0(1)$ is used to update the initial condition (e.g. $\widehat{z_1(0)}$) so that the end-time value $\overline{z_1(1)}$ is closer to target. This process is repeated until convergence. Besides the initial condition, the shooting method can be applied to update other parameters.
\subsubsection{Multiple-shooting method} The multiple-shooting method \cite{bock1984multiple} is an extension of the shooting method to long time-series; it splits a long time-series into chunks, and applies the shooting method to each chunk. Integration of a dynamical system for a long time is typically subject to noise and numerical error, while solving short time-series is generally easier and more robust.

As shown in the right subfigure of Fig.~\ref{fig:shoot}, a guess of initial condition at time $t_0$ is denoted as $\widehat{z_0}$, and we can use any ODE solver to get the estimated integral curve $\overline{z(t)}, t\in [t_0, t_1]$. Similarly, we can guess the initial condition at time $t_1$ as $\widehat{z_1}$, and get $\overline{z(t)}, t\in [t_1, t_2]$ by integration as in Eq.~\ref{eq:estimate}. Note that each time chunk is shorter than the entire chunk ($\vert t_{i+1} - t_i \vert <\vert t_3-t_0 \vert, i \in \{ 1,2 \}$), hence easier to solve. The split causes another issue: the guess might not match estimation at boundary points (e.g. $\overline{z(t_1)} \neq \widehat{z_1}, \overline{z(t_2)} \neq \widehat{z_2}$). Therefore, we need to consider this error of mismatch when updating parameters, and minimizing this mismatch error is typically easier compared to directly analyzing the entire sequence.

The multiple-shooting method can be written as:
\begin{equation}
\label{eq:multi-shoot-loss}
    \operatorname*{argmin}_{\eta, z_0, ... z_N}J = \operatorname*{argmin}_{\eta, z_0, ... z_N} \sum_{i=0}^N \int_{t_i}^{t_{i+1}} \Big(\overline{z(\tau)} - \widetilde{z(\tau)} \Big)^2 \mathrm{d}\tau + \alpha \sum_{i=0}^N \Big( \overline{z(t_i)} - \widehat{z_i} \Big)^2
\end{equation}
\begin{equation}
\label{eq:estimate}
    \overline{z(t)} = \widehat{z_i} + \int_{t_i}^t f_\eta \big( \overline{ z(\tau)}, \tau \big) \mathrm{d} \tau,\ \ \ \ t_i < t < t_{i+1},\ \ \ i \in \{0, 1, 2,...N \}
\end{equation}
where $N$ is the total number of chunks discretized at points $\{t_0, ... t_N\}$, with corresponding guesses $\{\widehat{z_0}, ... \widehat{z_N}\}$. We use $\overline{z(t)}$ to denote the estimated curve as in Eq.~\ref{eq:estimate}; suppose $t$ falls into the chunk $[t_i,t_{i+1}]$, $z(t)$ is determined by solving the ODE from $(\widehat{z_i},t_i)$, where $\widehat{z_i}$ is the guess of initial state at $t_i$. We use $\widetilde{z(t)}$ to denote the observation. The first part in Eq.~\ref{eq:multi-shoot-loss} corresponds to the difference between estimation $\overline{z(t)}$ and observation $\widetilde{z(t)}$, while the second part corresponds to the mismatch between estimation (orange square, $\overline{z(t_i)}$) and guess (blue circle, $\widehat{z_i}$) at split time points $t_i$. The second part is weighted by a hyper-parameter $\alpha$. The ODE function $f$ is parameterized by $\eta$. The optimization goal is to find the best $\eta$ that minimizes loss in Eq.~\ref{eq:multi-shoot-loss}, besides model parameters $\eta$, we also need to optimize the guess $\widehat{z_i}$ for state at time $t_i, i \in \{0,1,...N\}$. 
Note that though previous work typically limits $f$ to have a linear form, we don't have such limitations. Instead, multiple-shooting is generic for general $f$. 
\subsection{Adjoint state method}
Our goal is to minimize the loss function in Eq.~\ref{eq:multi-shoot-loss}. Let $\theta = [\eta, z_0, ...,z_N]$ represent all learnable parameters. After fitting an ODE, we derive the gradient of loss $L$ $w.r.t$ parameter $\theta$ and state guess $\widehat{z_i}$ for optimization.
\subsubsection{Adjoint state equation} Note that different from discrete case, the gradient in continuous case is slightly complicated. We refer to the adjoint method \cite{pontryagin2018mathematical,zhuang2020adaptive,chen2018neural}. Consider the following problem:
\begin{equation}
\label{eq:adjoint_ode}
    \frac{d \overline{z(t)}}{dt}= f_\theta \Big( \overline{z(t)},t \Big),\ \ s.t.\ \ \overline{z(0)}=x,\ \ t \in [0,T], \ \ \theta=[\eta, z_0, ... z_N]
\end{equation}
\begin{equation}
\label{eq:adjoint:loss}
    \hat{y} = \overline{z(T)},\ \ J \Big(\hat{y},y\Big) = J \Big( \overline{z(0)}+\int_0^T f_\theta( \overline{z},t)dt, y \Big)
\end{equation}
where the initial condition $z(0)$ is specified by input $x$, output $\hat{y}= \overline{ z(T)}$. The loss function $J$ is applied on $\hat{y}$, with target $y$. Compared with Eq.~\ref{eq:formulation} to Eq.~\ref{eq:estimate}, for simplicity, we use $\theta$ to denote both model parameter $\eta$ and guess of initial conditions $\{\widehat{z_i}\}$. The Lagrangian is
\begin{equation}
    L = J \Big( \overline{z(T)}, y \Big) + \int_0^T \lambda(t)^ \top \Big[\frac{d \overline{z(t)}}{dt} - f_\theta( \overline{z(t)},t) \Big] dt
    \label{eq:loss}
\end{equation}
where $\lambda(t)$ is the continuous Lagrangian multiplier. Then we have the following:
\begin{equation}
    \frac{\partial J}{\partial \overline{z(T)}} + \lambda(T) = 0
    \label{eq:lambda_bound}
\end{equation}
\begin{equation}
    \frac{d\lambda(t)}{dt} +  \Big ( \frac{\partial f_\theta( \overline{z(t)},t)}{\partial \overline{z(t)}} \Big )^\top \lambda(t) = 0\ \ \forall t\in (0,T)
    \label{eq:lambda_ode}
\end{equation}
\begin{equation}
    \frac{dL}{d\theta} = \int_T^0 \lambda(t)^\top \frac{\partial f_\theta( \overline{ z(t)},t)}{\partial \theta}dt
\label{eq:analytic_grad}
\end{equation}
We skip the proof for simplicity. In general, the adjoint method determines the initial condition $\lambda(T)$ by Eq.~\ref{eq:lambda_bound}, then solves Eq.~\ref{eq:lambda_ode} to get the trajectory of $\lambda(t)$, and finally integrates $\lambda(t)$ as in Eq.~\ref{eq:analytic_grad} to get the final gradient. Note that Eq.~\ref{eq:lambda_bound} to Eq.~\ref{eq:analytic_grad} is generic for general $\theta$, and in case of Eq.~\ref{eq:multi-shoot-loss} and Eq.~\ref{eq:estimate}, we have $\theta = [ \eta, z_0, ... z_N ]$, and $ \nabla \theta = [ \frac{\partial L}{\partial \eta}, \frac{\partial L}{\partial z_0}, ... \frac{\partial L}{\partial z_N}]$. Note that we need to calculate $\frac{\partial f}{\partial z}$ and $\frac{\partial f}{\partial \theta}$, which can be easily computed by a single backward pass; we only need to specify the forward model without worrying about the backward, because automatic differentiation is supported in frameworks such as PyTorch and Tensorflow. After deriving the gradient of all parameters, we can update these parameters by general gradient descent methods.

Note that though $J(\overline{z(T)},y)$ is defined on a single point in Eq.~\ref{eq:loss}, it can be defined as the integral form in Eq.~\ref{eq:multi-shoot-loss}, or a sum of single-point loss and integral form. The key observation is that for any loss in the integral form, e.g. $\int_{t=0}^T loss(t) dt $, we can defined an auxiliary variable $F$ such that $\frac{d F(t)}{dt}=loss(t), F(0)=0$, then $F(T)$ is just the value of the integral; in this way, we can transform integral form $\int_{0}^T loss(t) dt$ into a single point form $F(T)$. 
\subsubsection{Adaptive checkpoint adjoint}
Eq.~\ref{eq:lambda_bound} to Eq.~\ref{eq:analytic_grad} are the analytical form of the gradient in the continuous case, yet the numerical implementation is crucial for empirical performance. Note that $\overline{z(t)}$ is solved in forward-time (0 to $T$), while $\lambda(t)$ is solved in reverse-time ($T$ to 0), yet the gradient in Eq.~\ref{eq:analytic_grad} requires both $\overline{z(t)}$ and $\lambda(t)$ in the integrand. Memorizing a continuous trajectory $\overline{z(t)}$ requires much memory; to save memory, most existing implementations forget the forward-time trajectory of $\overline{z(t)}$, and instead only record the end-time state $\overline{z(T)}$ and $\lambda(T)$ and solve Eq.~\ref{eq:adjoint_ode} and Eq.~\ref{eq:lambda_bound} to Eq.~\ref{eq:analytic_grad} in reverse-time on-the-fly. 

While memory cost is low, existing implementations of the adjoint method typically suffer from numerical error: since the forward-time trajectory (denoted as $\overrightarrow{z(t)}=\overline{z(t)}$) is deleted, and the reverse-time trajectory (denoted as $\overleftarrow{z(t)}$) is reconstructed from the end-time state $z(T)$ by solving Eq.~\ref{eq:adjoint_ode} in reverse-time, $\overrightarrow{z(t)}$ and $\overleftarrow{z(t)}$ cannot accurately overlap due to inevitable errors with numerical ODE solvers. The error $\overrightarrow{z(t)}-\overleftarrow{z(t)}$ propagates to the gradient in Eq.~\ref{eq:analytic_grad} in the $\frac{\partial f(z, t)}{\partial z}$ term. Please see \cite{zhuang2020adaptive} for a detailed explanation.

To solve this issue, the adaptive checkpoint adjoint (ACA) \cite{zhuang2020adaptive} records $\overrightarrow{z(t)}$ using a memory-efficient method to guarantee numerical accuracy. In this work, we use ACA for its accuracy.
\begin{algorithm}[t]
\SetAlgorithmName{Algorithm}{} \\
\textbf{Input} Observation $\widetilde{z(t)}$, number of chunks $N$, learning rate $lr$.\\
\textbf{Initialize} model parameter $\eta$, state $\{\widehat{z_i}\}_{i=0}^N$ at discretized points $\{t_i\}_{i=0}^N$ \\
\textbf{Repeat until convergence} \\
\hspace{8mm}(1) Estimate trajectory $\overline{z(t)}$ from current parameters by the multiple shooting method as in Eq.~\ref{eq:estimate}. \\ 
\hspace{8mm}(2) Compute the loss $J$ in Eq.~\ref{eq:multi-shoot-loss}, plug $J$ in Eq.~\ref{eq:loss}. Derive the gradient by the adjoint method as in Eq.~\ref{eq:lambda_bound} to Eq.~\ref{eq:analytic_grad}. \\
\hspace{8mm}(3) Update parameters $\theta \leftarrow \theta - lr \times \nabla \theta$
\caption{Multiple-shooting adjoint method}
\label{algo:MSA}
\end{algorithm}
\vspace{-1mm}
\subsection{Multiple-Shooting Adjoint (MSA) method}
\subsubsection{Procedure of MSA}MSA is a combination of the multiple-shooting and the adjoint method, which is 
generic for various $f$. Details are summarized in Algo.~\ref{algo:MSA}. MSA iterates over the following steps until convergence: (1) estimate the trajectory based on the current parameters, using the multiple-shoot method for integration; (2) compute the loss and derive the gradient using the adjoint method; (3) update the parameters based on the gradient.
\vspace{-1mm}
\subsubsection{Advantages of MSA} Previous work has used the multiple-shooting method for parameter estimation in ODEs \cite{peifer2007parameter}, yet MSA is different in the following aspects: (A) Suppose the parameters have $k$ dimensions. MSA uses an element-wise update, hence has only $O(k)$ computational cost in each step; yet the method in \cite{peifer2007parameter} requires the inversion of a $k \times k$ matrix, hence might be infeasible for large-scale systems. (B) The implementation of \cite{peifer2007parameter} does not tackle the mismatch between forward-time and reverse-time trajectory, while we use ACA \cite{zhuang2020adaptive} for accurate gradient estimation in step (2) of Algo.~\ref{algo:MSA}. (C) From a practical perspective, our implementation is based on PyTorch which supports automatic-differentiation, therefore we only need to specify the forward model $f$ without the need to manually compute the gradient $\frac{\partial f}{\partial z}$ and $\frac{\partial f}{\partial \theta}$. Hence, our method is off-the-shelf for general models, while the method of \cite{peifer2007parameter} needs to re-implement $\frac{\partial f}{\partial z}$ and $\frac{\partial f}{\partial \theta}$ for different $f$, and conventional DCM with EM needs to re-derive the entire algorithm when $f$ changes.

\subsection{Dynamic causal modeling}
We briefly introduce the dynamical causal modeling here. Suppose there are $p$ nodes (ROIs) and denote the observed fMRI time-series signal as $s(t)$, which is a $p$-dimensional vector at each time $t$. Denote the hidden neuronal state as $z(t)$;  then $z(t)$ and $s(t)$ are  $p$-dimensional vectors for each time point $t$. Denote the hemodynamic response function (HRF) \cite{lindquist2009modeling} as $h(t)$, and denote the external stimulation as $u(t)$, which is an $n$-dimensional vector for each $t$. The forward-model is:
\begin{equation}
\label{eq:dcm}
f\Big( [z(t)\ \  D(t)] \Big)=
\begin{bmatrix}
 d z(t) /dt \\
 d D(t)/dt
\end{bmatrix}
=
\begin{bmatrix}
 D(t) z(t) + C u(t) \\
 B u(t)
\end{bmatrix},\ \ D(0)=A
\end{equation}
\begin{equation}
    s(t) = \Big( z(t) + \epsilon(t) \Big) * h(t),\ \ \widetilde{z(t)} = z(t) + \epsilon(t) = Deconv\Big(s(t), h(t) \Big)
\end{equation}
where $\epsilon(t)$ is the noise at time $t$, which is assumed to follow an independent Gaussian distribution, and $*$ represents convolution operation. Note that a more general model would be $s(t)=\Big( z(t) + \epsilon_1(t) \Big) * h(t) + \epsilon_2 (t)$, where $\epsilon_1 (t)$ is the inherent noise in neuronal state $z(t)$, and $\epsilon_2(t)$ is the measurement noise. We omit $\epsilon_2 (t)$ for simplicity in this project; it's possible to model both noises, even model HRF as learnable parameters, but would cause a more complicated model and require more data for accurate parameter estimation.

\begin{SCfigure}[][t]
\includegraphics[width=0.4\textwidth]{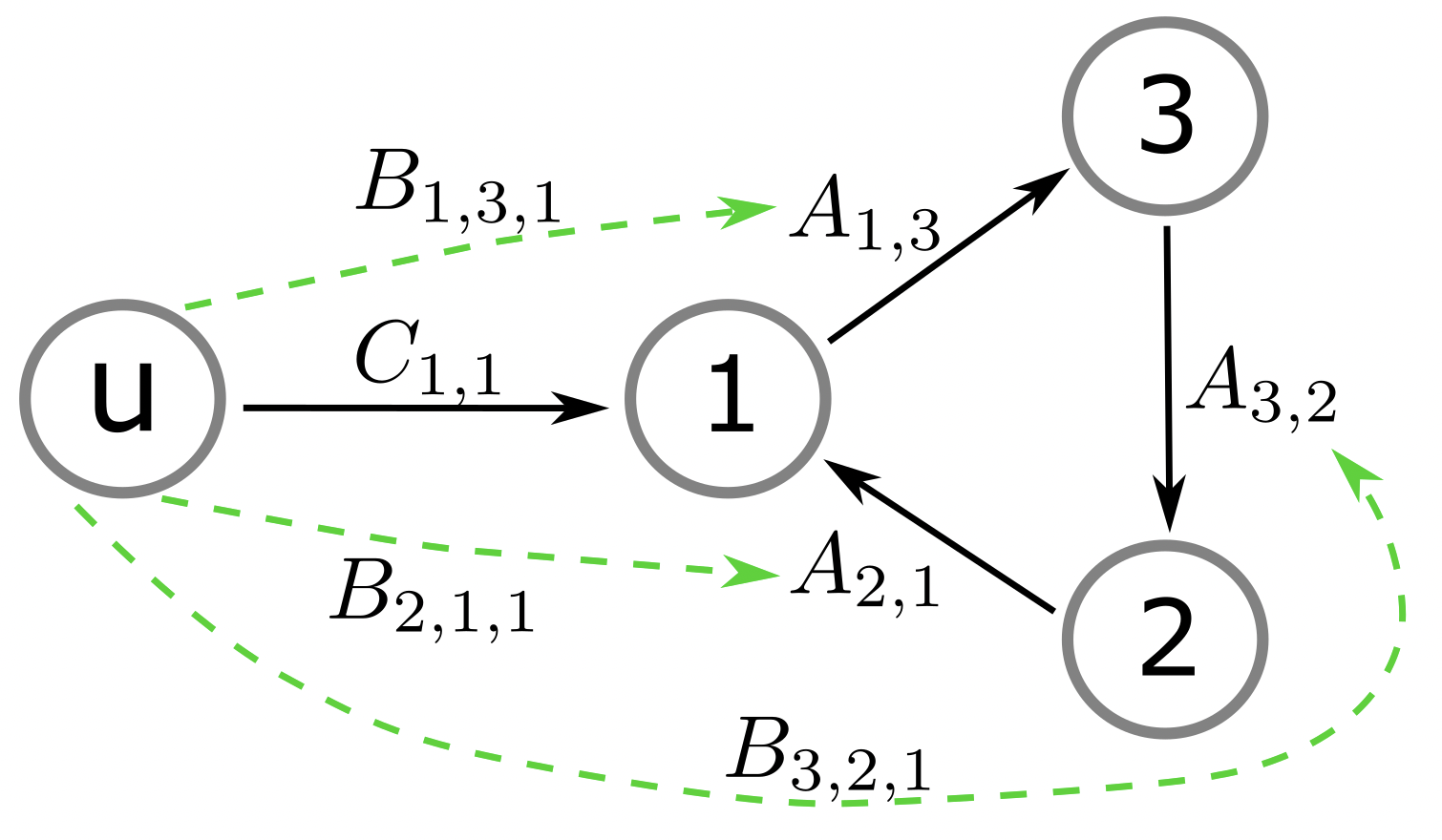}
\captionof{figure}{\small Toy example of dynamic causal modeling with 3 nodes (labeled 1 to 3). $u$ is a 1-D stimulation signal, so $n=1,p=3$. $A,B,C$ are defined as in Eq.~\ref{eq:dcm}. For simplicity, though $A$ is a $3\times3$ matrix, we assume only three elements $A_{1,3},A_{3,2},A_{2,1}$ are non-zero.}
\label{fig:toy}
\end{SCfigure}

$D(t)$ is a $p \times p$ matrix for each $t$, representing the effective connectome between nodes. $A$ is a matrix of shape $p \times p$, representing the interaction between ROIs. 
$B$ is a tensor of shape $p\times p \times n$, representing the effect of stimulation on the effective connectome. $C$ is a matrix of shape $p \times n$, representing the effect of stimulation on neuronal state. An example of $n=1,p=3$ is shown in Fig.~\ref{fig:toy}.

The task is to estimate parameters $A,B,C$ from noisy observation $s(t)$. For simplicity, we assume $h(t)$ is fixed and use the empirical result from Nitime project \cite{rokem2009nitime} in our experiments. By deconvolution of $s(t)$ with $h(t)$, we get a noisy observation of $z(t)$, denoted as $\widetilde{z(t)}$; $z(t)$ follows the ODE defined in Eq.~\ref{eq:dcm}. By plugging $f$ into Eq.~\ref{eq:formulation}, and viewing $\eta$ as $[A,B,C]$, this problem turns into a parameter estimation problem for ODEs, which can be efficiently solved by Algo.~\ref{algo:MSA}. We emphasize that Algo.~\ref{algo:MSA} is generic and in fact MSA can be applied to any form of $f$, where here the linear form of $f$ in Eq.~\ref{eq:dcm} is a special case for a specific model for fMRI. 

\section{Experiments}
\subsection{Validation on toy examples}
We first validate MSA on toy examples of linear dynamical systems, then validate its performance on large-scale systems and non-linear dynamical systems.

\begin{figure}
\begin{subfigure}{0.50\textwidth}
\includegraphics[width=\linewidth]{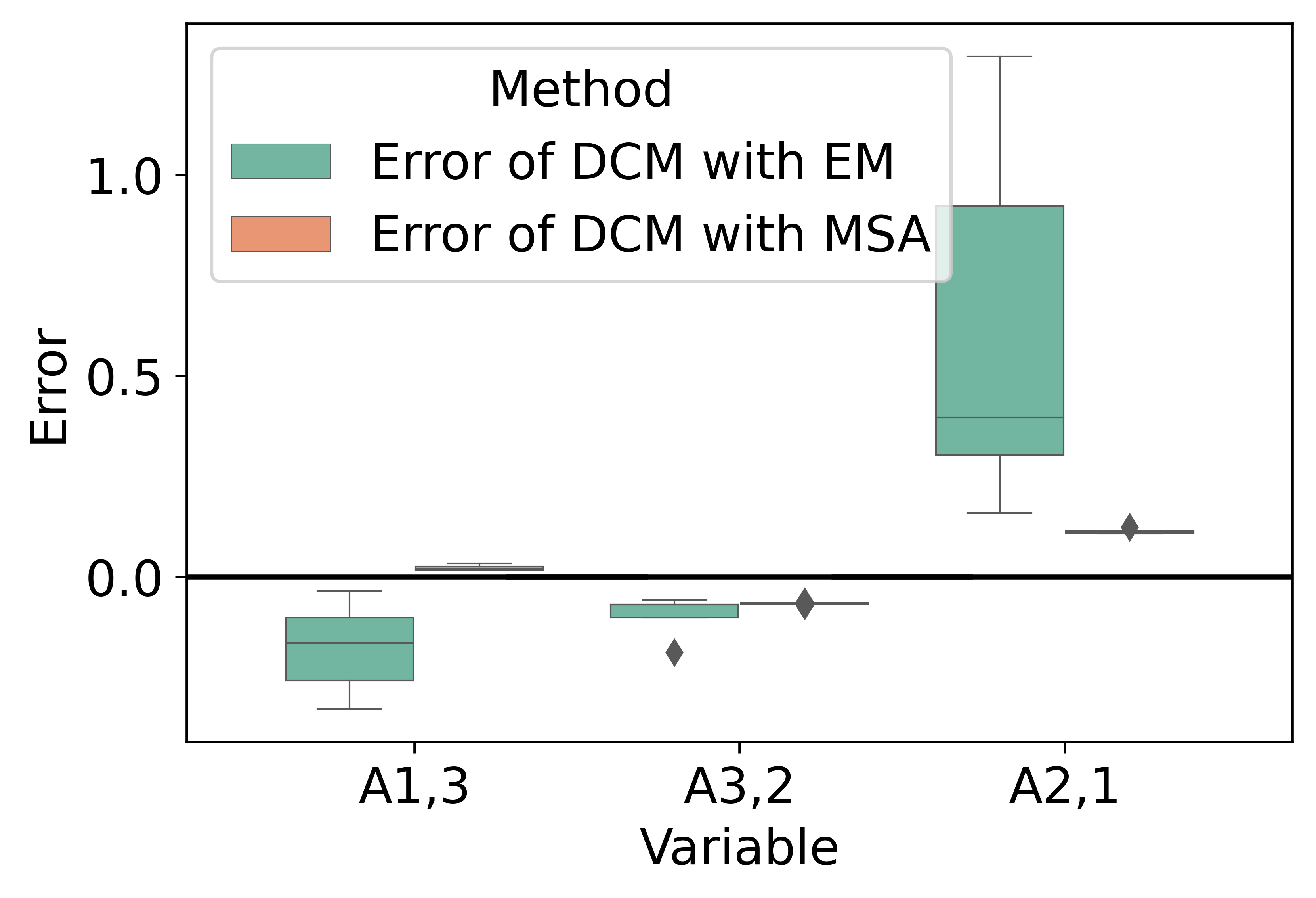}
\end{subfigure}
\begin{subfigure}{0.48\textwidth}
\includegraphics[width=\linewidth]{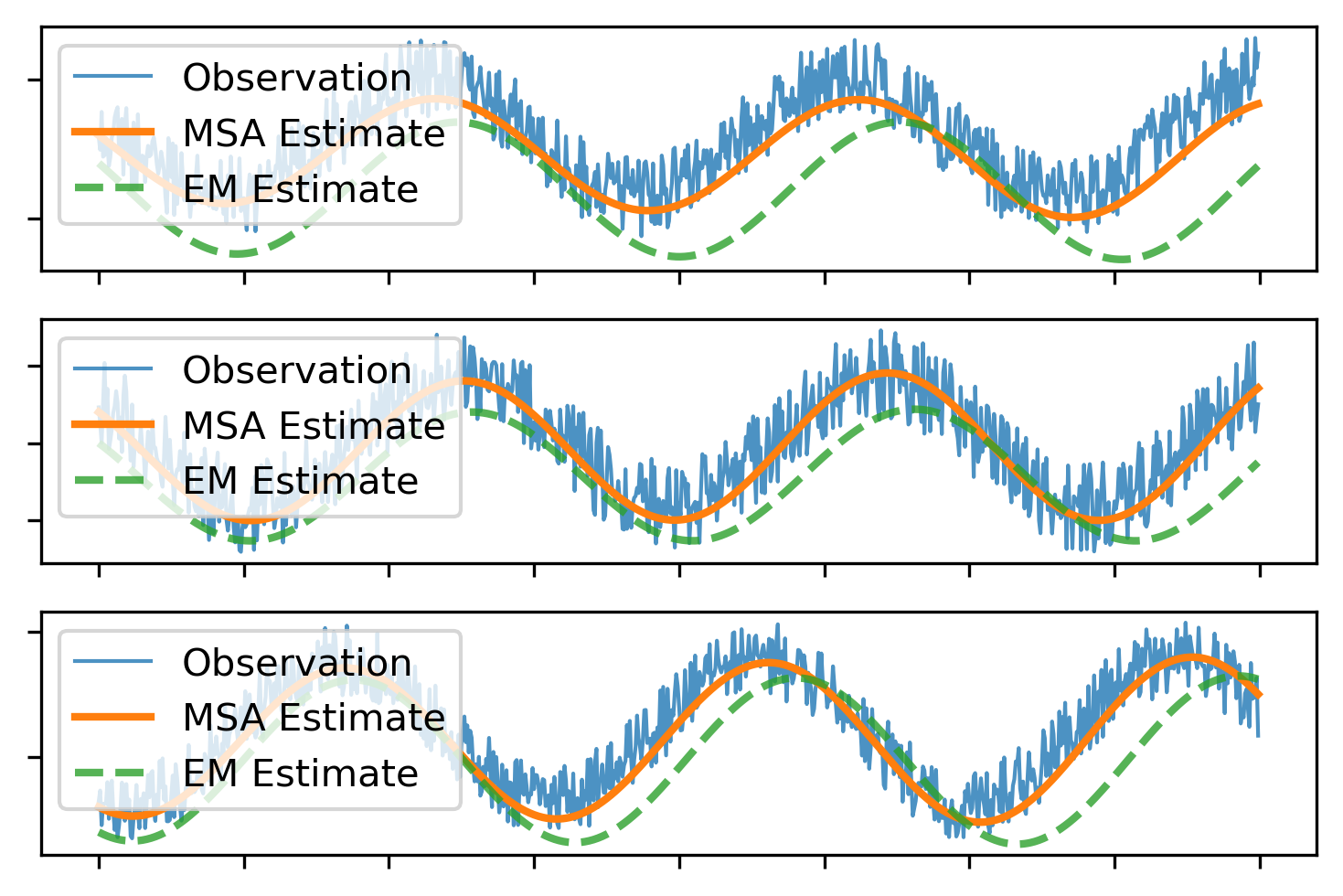}
\end{subfigure}
\caption{\small Results for the toy example of a linear dynamical system in Fig.~\ref{fig:toy}. Left: error in estimated value of connection $A_{1,3},A_{3,2},A_{2,1}$, other parameters are set as 0 in simulation. Right: from top to bottom are the results for node 1, 2, 3 respectively. For each node, we plot the observation and estimated curve from MSA and EM methods. Note that the estimated curve is generated by integration of the ODE under estimated parameters with only the initial condition known, not smoothing of 
noisy observation.}
\label{fig:dcm_error}
\end{figure}

\subsubsection{A linear dynamical system with 3 nodes} We first start with a simple linear dynamical system with only 3 nodes. We further simplify the matrix $A$ as in Fig.~\ref{fig:toy}, where only three elements in $A$ are non-zero. We set $B$ as a zeros matrix, and $u(t)$ as a 1-dimensional signal. The dynamical system is linear:
\begin{equation}
\label{eq:toy_small_linear}
\begin{bmatrix}
 d z(t)/dt \\
 d D(t)/dt
\end{bmatrix}
=
\begin{bmatrix}
 D(t) z(t) + C u(t) \\
 0
\end{bmatrix},\ D(0)=A,\ \  
u(t) = 
    \begin{cases}
    1, & floor(\frac{t}{2})\%2=0 \\
    0, & otherwise
    \end{cases}
\end{equation}
\begin{equation}
    \widetilde{z(t)} = z(t) + \epsilon(t),\ \ \ \epsilon(t) \sim N(0, \sigma^2)
\end{equation}
$u(t)$ is an alternating block function at a period of 2, taking values 0 or 1. The observed function $\widetilde{z(t)}$ suffers from $i.i.d$ Gaussian noise $\epsilon(t)$ with 0 mean and uniform variance $\sigma^2$.

We perform 10 independent simulations and parameter estimations. For estimation of DCM with the EM algorithm, we use the SPM package \cite{penny2011statistical}, which is a widely used standard baseline. The estimation in MSA is implemented in PyTorch, using ACA \cite{zhuang2020adaptive} as the ODE solver. For MSA, we use the AdaBelief optimizer \cite{zhuang2020adabelief} to update parameters with the gradient; though other optimizers such as SGD can be used, we found AdaBelief converges faster in practice.

For each of the non-zero elements in $A$, we show the boxplot of error in estimation in Fig.~\ref{fig:dcm_error}. Compared with EM, the error by MSA is significantly closer to 0 and has a smaller variance. An example of a noisy observation and estimated curves are shown in Fig.~\ref{fig:dcm_error}, and the estimation by MSA is visually closer to the ground-truth compared to the EM algorithm. We emphasize that the estimated curve is not a simple smoothing of the noisy observation; instead, after estimating the parameters of the ODE, the estimated curve (for $t>0$) is generated by solving the ODE using only the initial state. Therefore, the match between estimated curve and observation demonstrates that our method learns the underlying dynamics of the system.
\subsubsection{Application to large-scale systems}
After validation on a small system with only 3 nodes, we validate MSA on large scale systems with more nodes. We use the same linear dynamical system as in Eq.~\ref{eq:toy_small_linear}, but with the node number $p$ ranging from 10 to 100. Note that the dimension of $A$ and $B$ grows at a rate of $O(p^2)$, and the EM algorithm estimates the covariance matrix of size $O(p^4)$, hence the memory for EM method grows extremely fast with $p$. For various settings, the ground truth parameter is randomly generated from a uniform distribution between -1 and 1, and the variance of measurement noise is set as $\sigma=0.5$. For each setting, we perform 5 independent runs, and report the mean squared error (MSE) between estimated parameter and ground truth.

As shown in Table~\ref{table:large-scale}, for small-size systems (number of nodes $<= 20$), MSA consistently generates a lower MSE than the EM algorithm. For large-scale systems, since the memory cost of the EM algorithm is $O(p^4)$, the algorithm quickly runs out-of-memory. On the other hand, the memory cost for MSA is $O(p^2)$ because it only uses the first-order gradient. Hence, MSA is suitable for large-scale systems such as in whole-brain fMRI analysis. 

\subsubsection{Application to general non-linear systems}
Since neither the multiple-shoot method nor the adjoint state method requires the ODE $f$ to be linear, our MSA can be applied to general non-linear systems. Furthermore, since our implementation is in PyTorch which supports automatic differentiation, we only need to specify $f$ when fitting different models, and the gradient will be calculated automatically. Therefore, MSA is an off-the-shelf method, and is suitable for general non-linear ODEs both in theory and implementation.

We validate MSA on the Lotka-Volterra (L-V) equations \cite{volterra1928variations}, a system of non-linear ODEs describing the dynamics of predator and prey populations. The L-V equation can be written as:
\begin{equation}
    f\Big( [z_1(t), z_2(t)] \Big) = 
    \begin{bmatrix}
     d z_1(t)/ dt \\
     d z_2(t)/ dt
    \end{bmatrix}
    = 
    \begin{bmatrix}
     \zeta z_1(t) - \beta z_1(t) z_2(t) \\
     \delta z_1(t) z_2(t) - \gamma z_2(t) \\
    \end{bmatrix},\ \ 
    \begin{bmatrix}
     \widetilde{z_1(t)} \\
     \widetilde{z_2(t)}
    \end{bmatrix}
    = 
    \begin{bmatrix}
     z_1(t) + \epsilon_1(t) \\
     z_2(t) + \epsilon_2(t)
    \end{bmatrix}
\end{equation}
where $\zeta, \beta, \delta, \gamma$ are parameters to estimate, $\widetilde{z(t)}$ is the noisy observation, and $\epsilon(t)$ is the independent noise. Note that there are non-linear terms $z_1(t)z_2(t)$ in the ODE, making EM derivation difficult. Furthermore, the EM method needs to explicitly derive the posterior mean, hence needs to be re-derived for every different $f$; while MSA is generic and hence does not require re-derivation. 

Besides the L-V model, we also consider a modified L-V model, defined as:
\begin{align}
    dz_1(t)/dt &= \zeta z_1(t) - \beta \phi( z_2(t) )z_1(t) z_2(t) \\
    dz_2(t)/dt &= \delta \phi(z_1(t)) z_1(t) z_2(t) - \gamma z_2(t)
\end{align}
where $\phi(x)=1/(1+e^{-x})$ is the sigmoid function. We use this example to demonstrate the ability of MSA to fit highly non-linear ODEs.

\begin{figure}[t]
\begin{minipage}[]{0.49\textwidth}
\includegraphics[width=\linewidth]{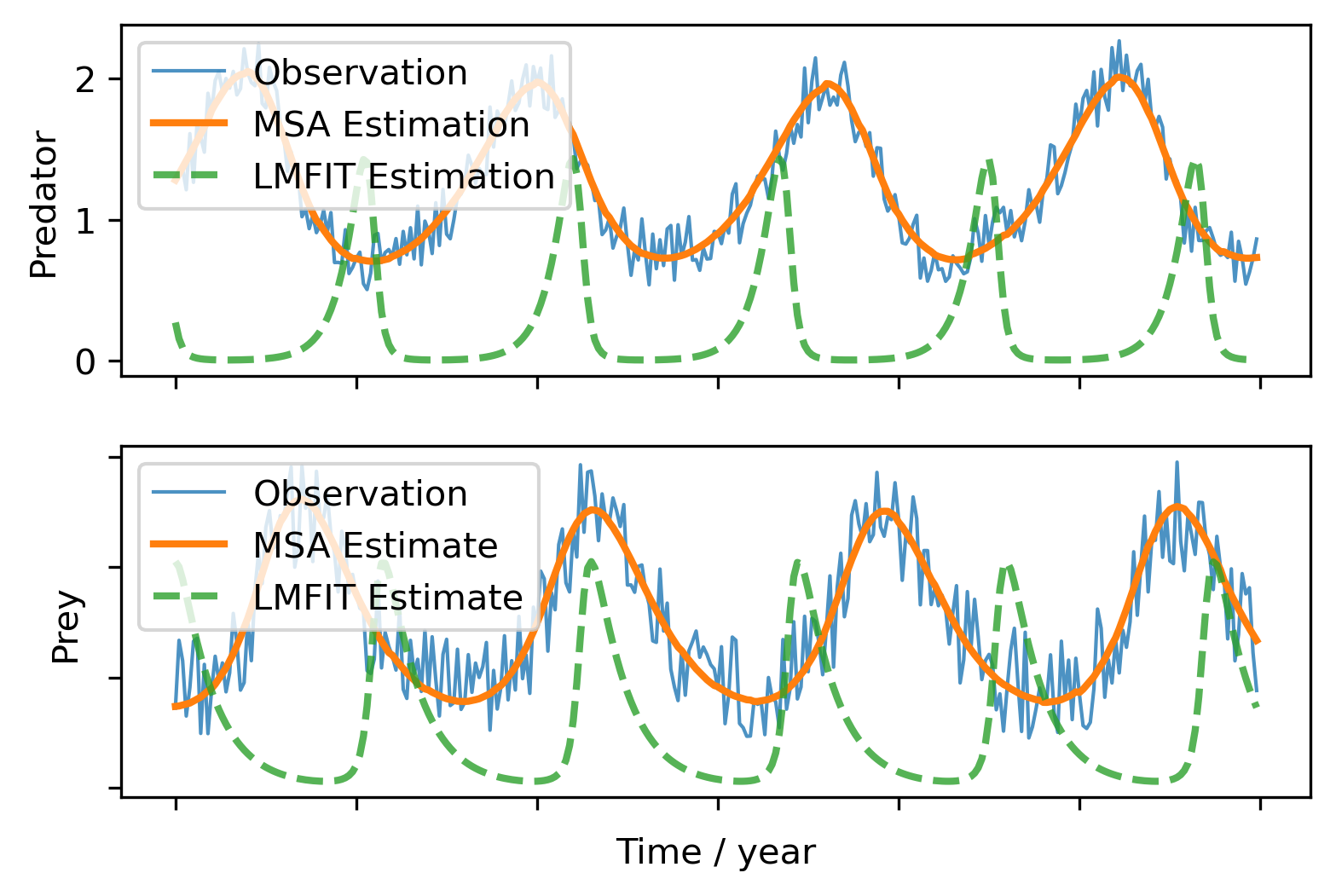}
\vspace{-4mm}
\captionof{figure}{\small Results for the L-V model.}
\label{fig:lotka}
\end{minipage}
\hfill
\begin{minipage}[]{0.49\textwidth}
\includegraphics[width=\linewidth]{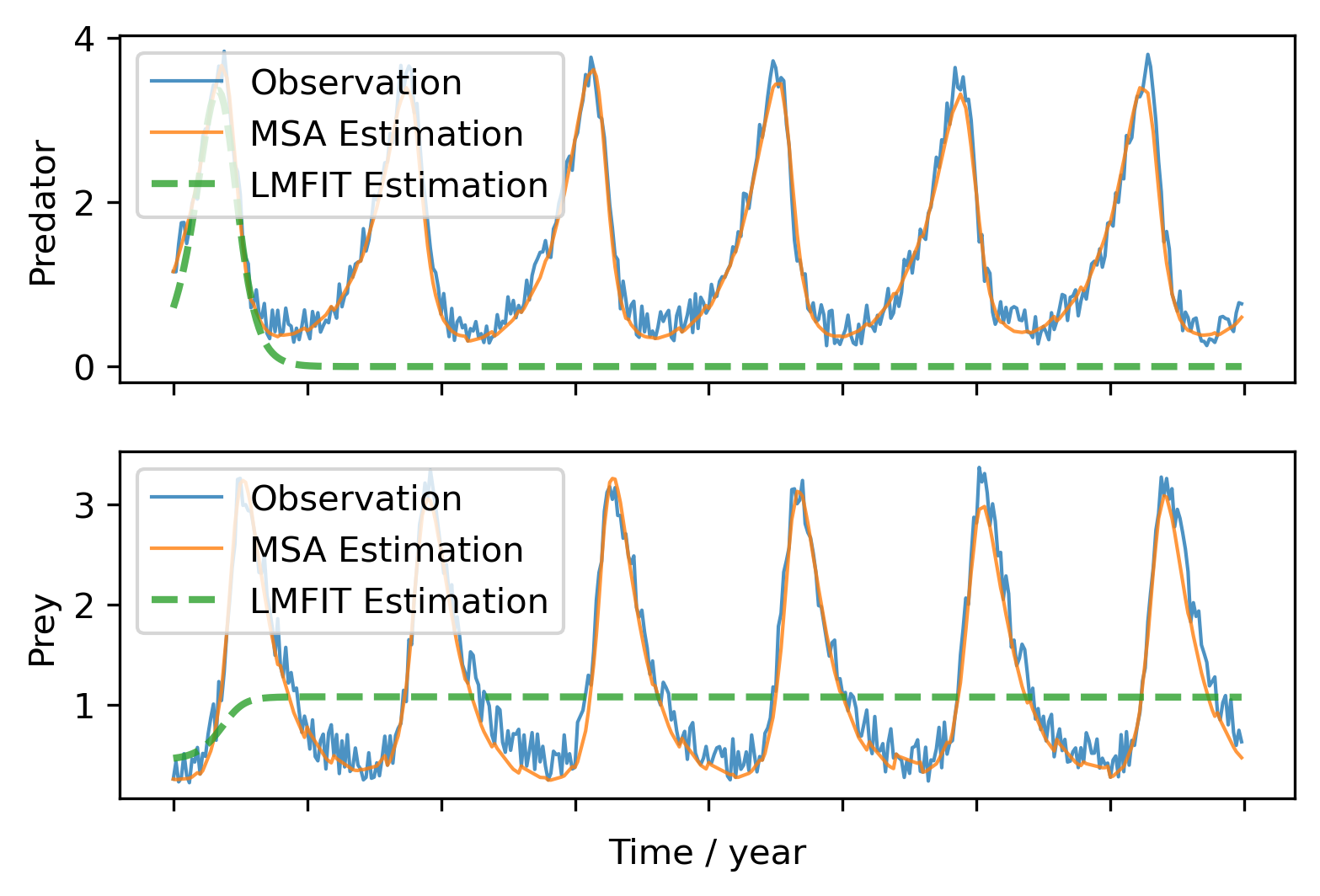}
\vspace{-4mm}
\captionof{figure}{\small Results for the modified L-V model.}
\label{fig:modify-lv}
\end{minipage}
\end{figure}

We compare MSA with LMFIT \cite{newville2016lmfit}, which is a well-known python package for non-linear fitting. We use L-BFGS solver in LMFIT, which generates better results than other solvers. We did not compare with original DCM with EM because it's unsuitable for general non-linear models. The estimation of the curve for $t>0$ is solved by integrating using the estimated parameters and initial conditions. As shown in Fig.~\ref{fig:lotka} and Fig.~\ref{fig:modify-lv}, compared with LMFIT, MSA recovers the system accurately. LMFIT directly fits the long sequences, while MSA splits long-sequences into chunks for robust estimation, which may partially explain the better performance of MSA.
\begin{table}[t]
\centering
\caption{\small Mean squared error ($\times 10^{-3}$, \textbf{lower} is better) in estimation of parameters for a linear dynamical system with different number of nodes. ``OOM'' represents ``out of memory''.}
\label{table:large-scale}
\begin{tabular}{c|cccc}
\hline
    & 10 Nodes                   & 20 Nodes                   & 50 Nodes                   & 100 Nodes                  \\ \hline
EM  & $3.3 \pm 0.2$ & $3.0 \pm 0.2$ & OOM                        & OOM                        \\
MSA & $0.7 \pm 0.1$ & $0.9 \pm 0.3$ & $0.8 \pm 0.1$ & $0.8 \pm 0.2$ \\ \hline
\end{tabular}
\end{table}


\begin{figure}[t]
\begin{subfigure}{0.30\textwidth}
\includegraphics[width=\linewidth]{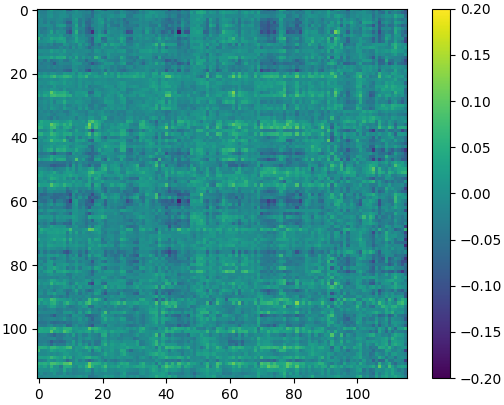}
\end{subfigure}
\begin{subfigure}{0.30\textwidth}
\includegraphics[width=\linewidth]{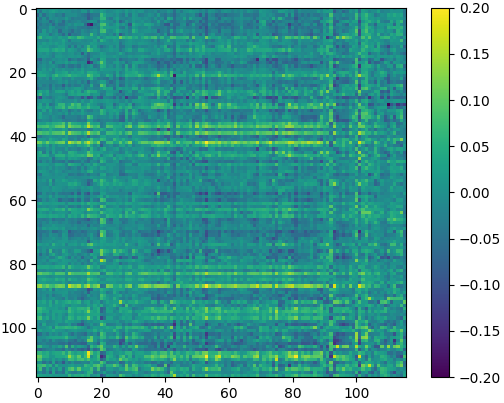}
\end{subfigure}
\begin{subfigure}{0.35\textwidth}
\includegraphics[width=\linewidth]{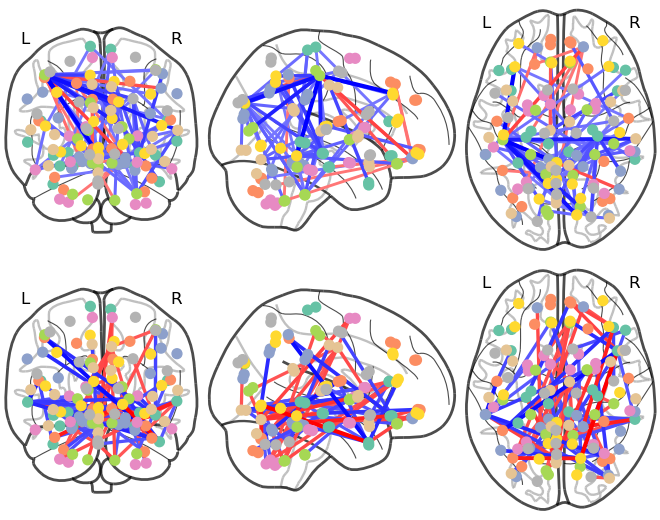}
\end{subfigure}
\caption{\small An example of MSA for one subject. Left: effective connectome during task 1. Middle: effective connectome during task 2. Right: top and bottom represents the effective connectome for task 1 and 2 respectively. Blue and red edges represent positive and negative connections respectively. Only top 5\% strongest connections are visualized.}
\label{fig:connectome}
\end{figure}

\begin{figure}[t]
\begin{subfigure}{0.31\textwidth}
\includegraphics[width=\linewidth]{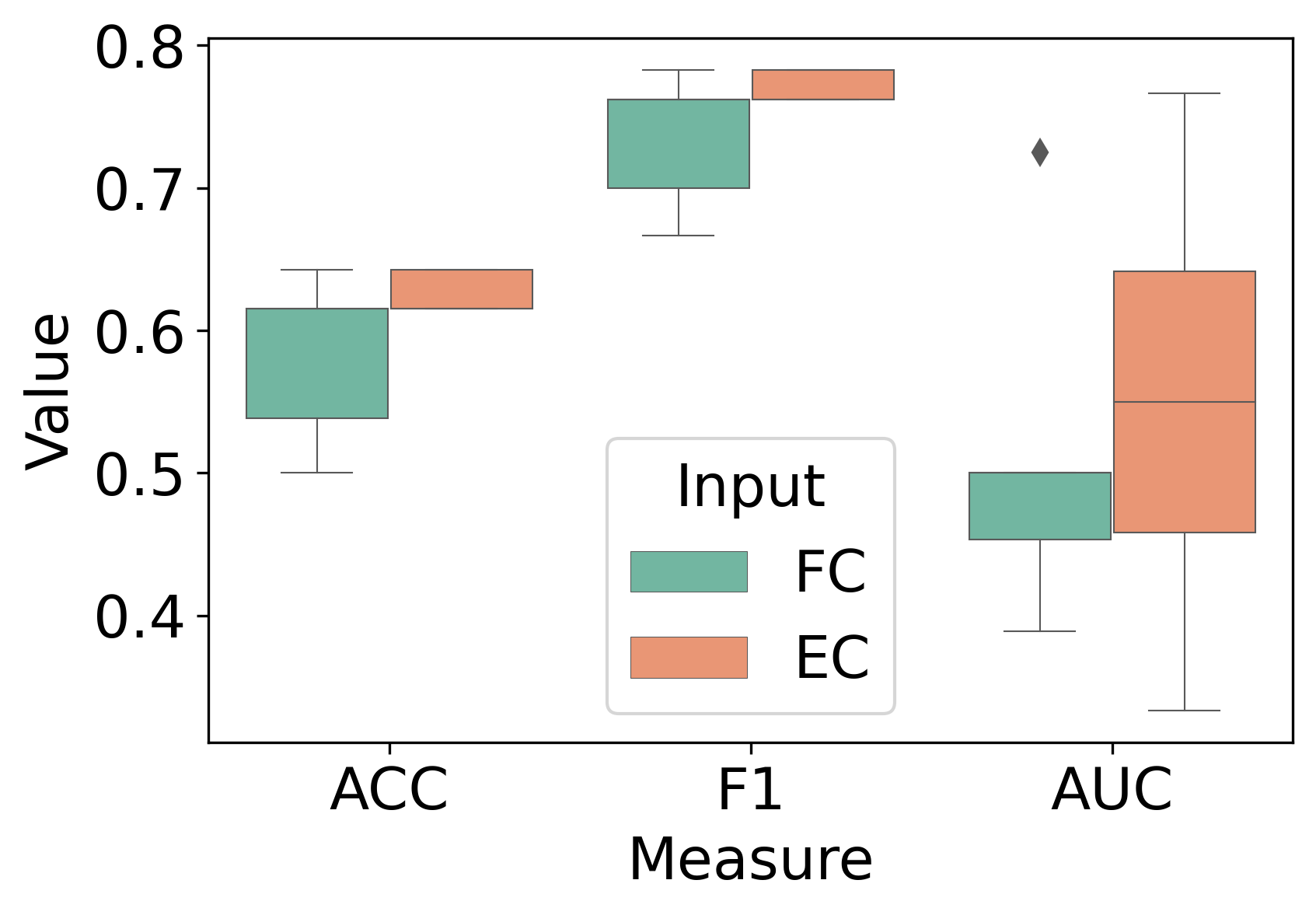}
\vspace{-3mm}
\caption{\small Classification result based on Random Forest.}
\end{subfigure}
\begin{subfigure}{0.32\textwidth}
\includegraphics[width=\linewidth]{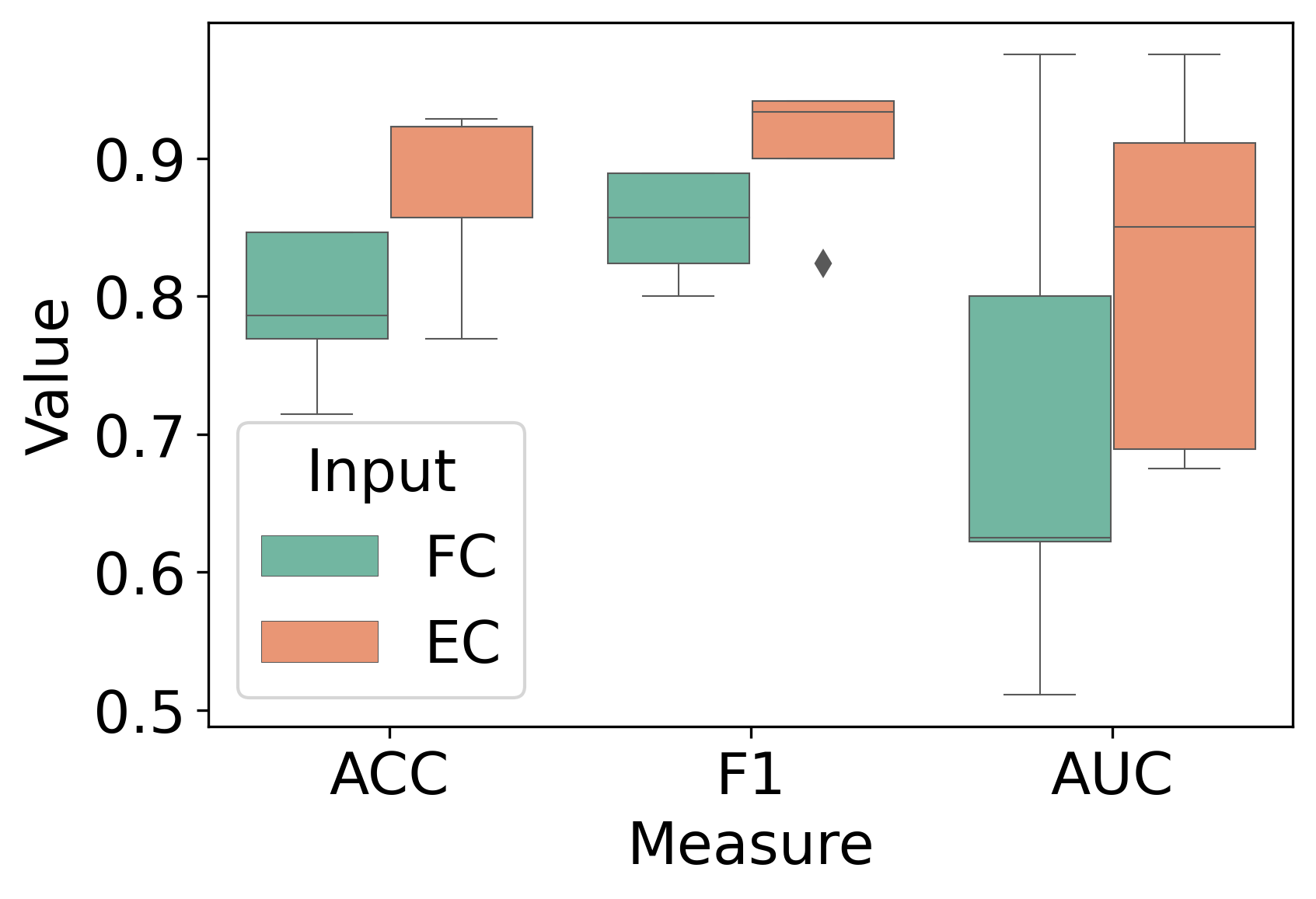}
\vspace{-3mm}
\caption{\small Classification result based on InvNet.}
\end{subfigure}
\begin{subfigure}{0.32\textwidth}
\includegraphics[width=\linewidth]{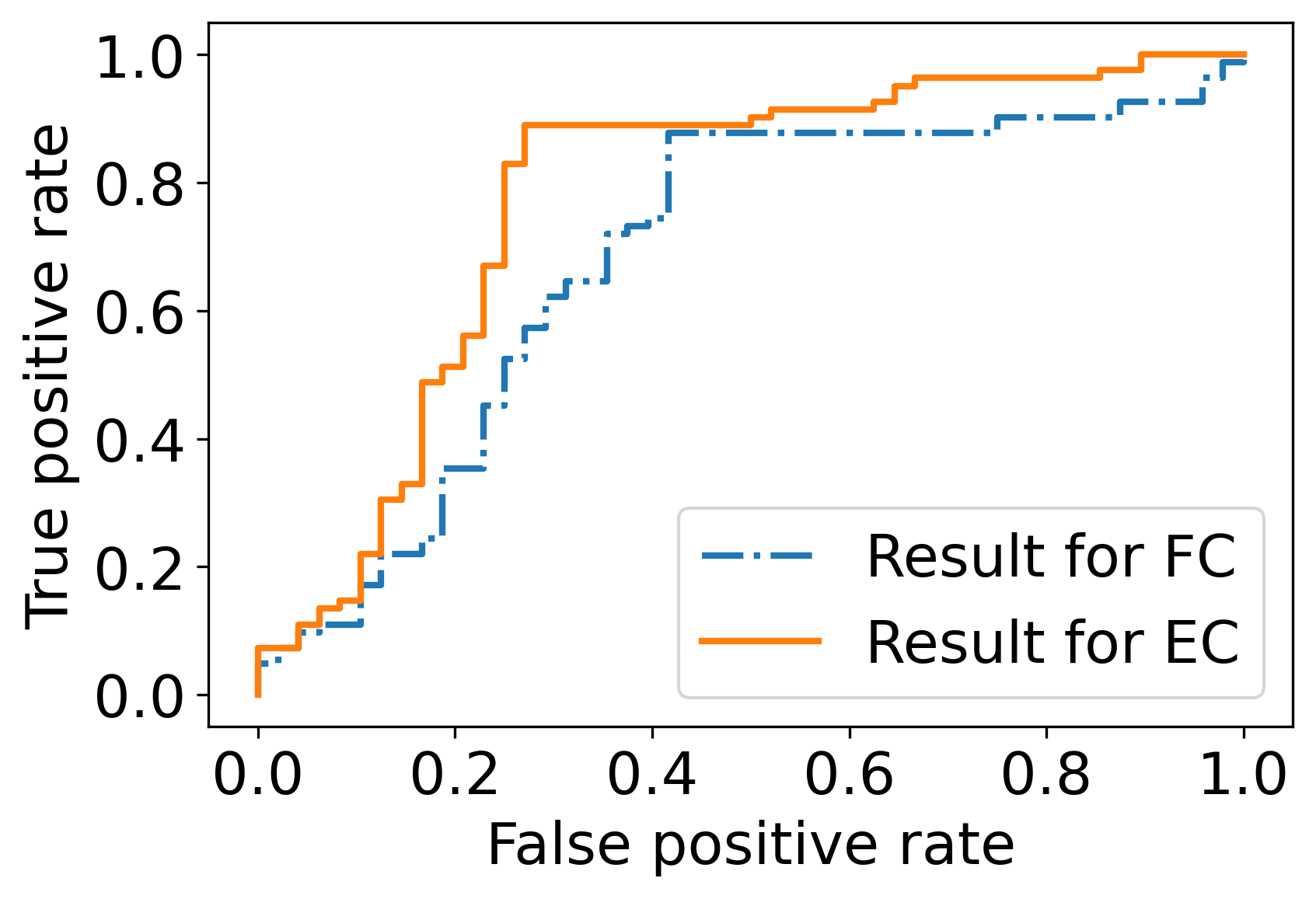}
\vspace{-3mm}
\caption{\small ROC-AUC curve for results with InvNet.}
\end{subfigure}
\vspace{-2mm}
\caption{\small Classification results for ASD vs. control.}
\label{fig:classification}
\end{figure}
\subsection{Application to whole-brain dynamic causal modeling with fMRI}
We apply MSA on  whole-brain fMRI analysis with dynamic causal modeling. fMRI for 82 children with ASD and 48 age and IQ-matched healthy controls were acquired. A biological motion perception task and a scrambled motion task \cite{kaiser2010neural} were presented in alternating blocks. The fMRI (BOLD, 132 volumes, TR = 2000ms, TE = 25ms, flip angle = 60$^{\circ}$, voxel size 3.44×3.44×4 $mm^3$) was acquired on a Siemens MAGNETOM Trio 3T scanner. 

\subsubsection{Estimation of EC}
We use the AAL atlas \cite{tzourio2002automated} containing 116 ROIs. For each subject, the parameters for dynamic causal modeling as in Eq.~\ref{eq:dcm} is estimated using MSA. An example snapshot of the effective connectome (EC) during the two tasks is shown in Fig.~\ref{fig:connectome}, showing MSA captures the dynamic EC during different tasks.

\subsubsection{Classification task}
We conduct classification experiments for ASD vs. control using EC and FC as input respectively. The EC estimated by MSA at each time point provides a data sample, and the classification of a subject is based on the majority vote of the predictions across all time points. The FC is computed using Pearson correlation. We experimented with a random forest model and InvNet \cite{zhuang2019invertible}. Results for a 10-fold subject-wise cross validation are shown in Fig.~\ref{fig:classification}. For both models, using EC as input generates better accuracy, F1 score and AUC score (threshold range is [0,1]). This indicates that estimating the underlying dynamics of fMRI helps identification of ASD.

\section{Conclusion}
We propose the multiple-shooting adjoint (MSA) method for parameter estimation in ODEs, enabling whole-brain dynamic causal modeling. MSA has the following advantages: robustness for noisy observations, ability to handle large-scale systems, and a general off-the-shelf framework for non-linear ODEs. We validate MSA in extensive toy examples and apply MSA to  whole-brain fMRI analysis with DCM. To our knowledge, our work is the first to successfully apply whole-brain dynamic causal modeling in a classification task based on fMRI. Finally, MSA is generic and can be applied to other problems such as EEG and modeling of biological processes.
\small
\bibliography{example_paper.bib}

\begin{thebibliography}{10}
\providecommand{\url}[1]{\texttt{#1}}
\providecommand{\urlprefix}{URL }
\providecommand{\doi}[1]{https://doi.org/#1}

\bibitem{bock1984multiple}
Bock, H.G., Plitt, K.J.: A multiple shooting algorithm for direct solution of
  optimal control problems. IFAC Proceedings Volumes  (1984)

\bibitem{chen2018neural}
Chen, R.T., Rubanova, Y., Bettencourt, J., Duvenaud, D.K.: Neural ordinary
  differential equations. Advances in neural information processing systems
  (2018)

\bibitem{di2017enhancing}
Di~Martino, A., O’connor, D., Chen, B., Alaerts, K., Anderson, J.S., et~al.:
  Enhancing studies of the connectome in autism using the autism brain imaging
  data exchange ii. Scientific data  (2017)

\bibitem{frassle2020regression}
Fr{\"a}ssle, S., Harrison, S.J., Heinzle, J., Clementz, B.A., Tamminga, C.A.,
  et~al.: Regression dynamic causal modeling for resting-state fmri. bioRxiv
  (2020)

\bibitem{friston2003dynamic}
Friston, K.J., Harrison, L.: Dynamic causal modelling. Neuroimage  (2003)

\bibitem{hildebrand1987introduction}
Hildebrand, F.B.: Introduction to numerical analysis (1987)

\bibitem{kaiser2010neural}
Kaiser, M.D., Hudac, C.M., Shultz, S., Lee, S.M., Cheung, C., et~al.: Neural
  signatures of autism. PNAS  (2010)

\bibitem{kiebel2008dynamic}
Kiebel, S.J., Garrido, M.I., Moran, R.J., Friston, K.J.: Dynamic causal
  modelling for eeg and meg. Cognitive neurodynamics  (2008)

\bibitem{lindquist2009modeling}
Lindquist, M.A., Loh, J.M., Atlas, L.Y., Wager, T.D.: Modeling the hemodynamic
  response function in fmri: efficiency, bias and mis-modeling. Neuroimage
  \textbf{45} (2009)

\bibitem{moon1996expectation}
Moon, T.K.: The expectation-maximization algorithm. ISPM  (1996)

\bibitem{nation2006patterns}
Nation, K., Clarke, P., Wright, B., Williams, C.: Patterns of reading ability
  in children with autism spectrum disorder. J Autism Dev Disord  (2006)

\bibitem{newville2016lmfit}
Newville, M., Stensitzki, T., Allen, D.B., Rawlik, M., Ingargiola, A., Nelson,
  A.: Lmfit: Non-linear least-square minimization and curve-fitting for python
  (2016)

\bibitem{peifer2007parameter}
Peifer, M., Timmer, J.: Parameter estimation in ordinary differential equations
  for biochemical processes using the method of multiple shooting  (2007)

\bibitem{penny2011statistical}
Penny, W.D., Friston, K.J., Ashburner, J.T., Kiebel, S.J., Nichols, T.E.:
  Statistical parametric mapping: the analysis of functional brain images
  (2011)

\bibitem{pontryagin2018mathematical}
Pontryagin, L.S.: Mathematical theory of optimal processes (2018)

\bibitem{prando2020sparse}
Prando, G., Zorzi, M., Bertoldo, A., Corbetta, M., Chiuso, A.: Sparse dcm for
  whole-brain effective connectivity from resting-state fmri data. NeuroImage
  (2020)

\bibitem{razi2017large}
Razi, A., Seghier, M.L., Zhou, Y., McColgan, P., Zeidman, P., Park, H.J.,
  et~al.: Large-scale dcms for resting-state fmri. Network Neuroscience  (2017)

\bibitem{rokem2009nitime}
Rokem, A., Trumpis, M., Perez, F.: Nitime: time-series analysis for
  neuroimaging data. In: Proceedings of the 8th Python in Science Conference
  (2009)

\bibitem{seghier2010identifying}
Seghier, M.L., Zeidman, P., Leff, A.P., Price, C.: Identifying abnormal
  connectivity in patients using dynamic causal modelling of fmri responses.
  Front. Neurosci  (2010)

\bibitem{tzourio2002automated}
Tzourio-Mazoyer, N., Landeau, B., Papathanassiou, D., Crivello, F., Etard, O.,
  et~al.: Automated anatomical labeling of activations in spm using a
  macroscopic anatomical parcellation of the mni mri single-subject brain.
  Neuroimage  (2002)

\bibitem{van2010exploring}
Van Den~Heuvel, M.P., Pol, H.E.H.: Exploring the brain network: a review on
  resting-state fmri functional connectivity. Eur Neuropsychopharmacol  (2010)

\bibitem{volterra1928variations}
Volterra, V.: Variations and fluctuations of the number of individuals in
  animal species living together. ICES Journal of Marine Science  \textbf{3}
  (1928)

\bibitem{zhuang2020adaptive}
Zhuang, J., Dvornek, N., Li, X., Tatikonda, S., Papademetris, X., Duncan, J.:
  Adaptive checkpoint adjoint for gradient estimation in neural ode. ICML
  (2020)

\bibitem{zhuang2019invertible}
Zhuang, J., Dvornek, N.C., Li, X., Ventola, P., Duncan, J.S.: Invertible
  network for classification and biomarker selection for asd. In: MICCAI (2019)

\bibitem{zhuang2020adabelief}
Zhuang, J., Tang, T., Ding, Y., Tatikonda, S.C., Dvornek, N., Papademetris, X.,
  Duncan, J.: Adabelief optimizer: Adapting stepsizes by the belief in observed
  gradients. NeurIPS  (2020)

\end{thebibliography}
\bibliographystyle{splncs04}

\end{document}